\documentclass[preprint]{aastex}
\usepackage{caption2}
\slugcomment{2020.12}

\shorttitle{Comments on the radiation-quiet anti-glitch} \shortauthors{H.Tong}

\begin{document}

\title{The dissipation of toroidal magnetic fields and spin-down evolution of young and strongly magnetized pulsars}

\author{Zhi-Fu, Gao\altaffilmark{1,2}, Hao Shan\altaffilmark{1}, Hui Wang\altaffilmark{3} }

\altaffiltext{1}{Xinjiang Astronomical Observatory, CAS,150, Science 1-Street, Urumqi, Xinjiang, 830011, China; }
\altaffiltext{2}{ Key Laboratory of Radio Astronomy, CAS, West Beijing Road, Nanjing, 210008, China  ;}
\altaffiltext{3}{ School of Physical Science and Technology, Southwest Jiaotong University, Chengdu, 610031, China  ;\\ Corresponding author: zhifugao@xao.ac.cn }

\begin{abstract}
Magnetars are a kind of pulsars powered mainly by superhigh magnetic fields. They are popular sources with many unsolved issues in themselves, but also linked to various high energy phenomena, such as QPOs, giant flares, fast radio bursts and super-luminous supernovae. In this work, we first review our recent works on the dissipation of toroidal magnetic fields in magnetars and rotationally powered pulsars, then review the spin-down evolution of young and strongly magnetized pulsars, especially of magnetars. We present an interesting and important relation between the magnetization parameter, and magnetic field in the magnetar crust. Finally, we introduce our two works in progress: to explain the magnetar "anti-gltich" events in the thermal plastic flow model and to revisit the expression of braking index $n$, which is independent of the second derivative of spin frequency of a pulsar and give some proposals for our future work.
\end{abstract}


\section{Introduction}
Megnetars are a kind of pulsars powered by their magnetic energy rather than their rotational energy. They are categorized into two populations historically: Soft Gamma$-$ray Repeaters\,(SGRs) and Anomalous X$-$ray Pulsars\,(AXPs)\,(Duncan \& Thompson 1992). Observations show that part of magnetars in quiescent state have soft X-ray radiations, and their typical values of soft X-ray luminosity $L_{X}$ are about $10^{34}-10^{36}$\,erg\,s$^{-1}$.

Beloborodov \& Li (2016) also believed that the soft X-ray luminosity of magnetars might be related to the Ohmic decay of the magnetic field, but they did not take into the effects of general relativity on Ohmic decay of toroidal magnetic fields. Wang et al. (2019) derived an eigenvalue equation for the Ohmic decay of toroidal magnetic field in the framework of general relativity. Gao et al. (20190 calculated the magnetic energy decay rates and soft X-ray luminosities for 22 magnetars. Recently, Kaspi \& Beloborodov (2017) discussed the relation between the ohmic decay and soft X-ray luminosity of magnetars. There have been a series of recent studies of magnetar activities from the crustal plastic flow, which occurs in hot crust with $T >10^{8}$\,(Kaspi and Beloborodov 2017). To be specific, Li et al.\,(2016) modeled magnetar outbursts from plastic failures triggered by thermoplastic waves and Hall-mediated avalanches in the crust that dissipates magnetic energy.

The spin-down evolution of pulsars has become an important research hot spot in the field of compact objects\,(Dan et al. 2020; Yan et al. 2011a, 2011b, 2018; Yuan et al. 2017). When the magneto-dipole radiation\,(MDR) solely causes the pulsar spin-down, the braking index is predicted to be $n=3$. Up to date, only 9 of the $\sim 3000$ known pulsars have reliable measured braking indices, all of which deviate from 3, demonstrating that the spin-down mechanism is not pure MDR. The present braking theories of pulsars are challenged by both of two relatively small and high braking indice of PSR J1734$-$3333 and PSR J1640$-$4631\,(Espinoza et al. 2011; Archibald et al. 2016) .

The reminder of this work is organized as follows. We review our works on the dissipation of toroidal magnetic fields of magnetars in Sec.\,2, review our works on the braking indices of strongly magnetized pulsars, including magnetars in Sec.\,3, introduce our current works in progress in Sec.\,(4), and give a summary and outlook in Sec.\,5.

\section{Toroidal magnetic field dissipation }
This Section includes two parts, in which we review the Ohmic decay in the crust and thermoplastic wave heating due to the dissipation of toroidal magnetic field, respectively.
\subsection{Ohmic decay in the crust}
By implementing realistic conductivity profiles provided by Potekhin et al.(2015) into the code (http://www.ioffe.ru/astro/conduct)and combine with the equation of state\,(EoS), we  calculated the NS inner crustal conductivity\,$\sigma$. Since the effect of a strong magnetic field on electrical conductivity was taking into account in Potekhin et al. (2015), our results will be more reliable and will better reflect the actual situation of the crustal conductivity. Firstly, we preferred a particular configuration of the force-free fields, satisfying
\begin{equation}
\nabla \times \vec{B}=\mu\vec{B},~~~~\vec{B}\cdot \nabla \mu=0,
\end{equation}
where $\mu$ is a parameter related to the magnetic field curvature and depends on the EoS.
For an axially symmetric field $\vec{B}$, it is convenient to decompose it into the so-called poloidal and toroidal parts by $\vec{B}_{p}=B_{r}\,\vec{e}_{r}+B_{\theta}\,\vec{e}_{\theta}$ and $\vec{B}_{t}=B_{\varphi}\,\vec{e}_{\varphi}$, where $\vec{e}_{r}$, $\vec{e}_{\theta}$ and $\vec{e}_{\varphi}$ represent unit vectors in the $r-$direction, $\theta-$direction and $\varphi-$direction, respectively.
There is a simple linear relation between $B_{\varphi}$ and $B_{r}$, $|B_{\varphi}|/|B_{r}|= \mu xR_{NS}/2$, where $R_{NS}=R/{\rm km}$ is the dimensionless NS radius. Thus, we have
\begin{equation}
B_{\varphi}=\frac{1}{2}\mu xRB\sum_{n}A_{n}\frac{j_{1}(n\pi x)}{xR^{2}}e^{\frac{-c^{2}\lambda_{n}t}{4\pi\sigma R^{2}}}sin\theta,
\end{equation}
where $x=r/R$, $j_{1}(\zeta)=\frac{1}{\zeta^{2}}(\sin\zeta-\zeta\cos\zeta)$ is the first-order spherical Bessel function, $A_{n}$ is the expansion coefficient $A_{n}=\int^{1}_{0}xj_{1}(n\pi x)x^{2}dx/(\int^{1}_{0}j_{1}^{2}(n\pi x)x^{2}dx)$, and
$\lambda_{n}$ is determined by the associated boundary-regularity conditions.
We used the TMA model(Singh \& Saxena 2012), and selected a magnetar with a typical mass of $M=1.45M_{\odot}$, corresponding to $R=11.77$ and $\mu=1.676$. From Wang et al.(2019), we obtained the decay rate of $B_{p}$,
\begin{equation}
\frac{dB_{p}}{dt}=\frac{dB_{r}}{dt}=B\frac{-c^{2}\lambda_{n}}{4\pi\sigma R}\Sigma_{n}A_{n}\frac{j_{1}(n\pi x)}{x^{2}R^{2}}e^{\frac{-c^{2}\lambda_{n}t}{4\pi\sigma R^{2}}}cos\theta
\end{equation}
and the decay rate of $B_{t}$
\begin{equation}
\frac{dB_{t}}{dt}=\frac{dB_{\varphi}}{dt}=-B\frac{\mu xc^{2}\lambda_{n}}{8\pi\sigma R}\Sigma_{n}A_{n}\frac{j_{1}(n\pi x)}{x^{2}R^{2}}e^{\frac{-c^{2}\lambda_{n}t}{4\pi\sigma R^{2}}}sin\theta.
\end{equation}
The magnetic energy decay rates are estimated as
\begin{equation}
L_{p/t}=\frac{-1}{4\pi}\int_{v} B_{p/t}\frac{dB_{p/t}}{dt}dV,
\end{equation}
where $dV=4\pi r^{2}dr, R_{crust}\sim0.98\,{\rm km}$. The total magnetic
energy decay rate is mainly dominated by the toroidal component $L_{B}=L_{p}+L_{t}\approx L_{t}$. Based on the above expressions, we calculated the dipolar toroidal magnetic field decay rates, and investigated the relation of  soft X-ray luminosity $L_{\rm X}$ and totational energy loss rate $L_{\rm rot}$  of 22 magnetars. It is found that, for magnetars with $L_{\rm X}> L_{\rm rot}$, the Ohmic decay of crustal toroidal magnetic fields can provide their observed soft X-ray luminosities and maintain high thermal temperatures\,(Gao et al. 2019).

The crustal magnetic field of a NS usually decays at different timescales in the forms of Hall drift and Ohmic
dissipation. The magnetization parameter, $\omega_{B}\tau$, is defined as the ratio of the Ohmic timescale $\tau_{\rm Ohm}$ to the Hall drift timescale $\tau_{\rm Hall}$, where $\omega_{B}$ is the electron gyro-frequency, and $\tau$ is the electron relaxation time. We calculated the crust conductivity of NSs and gave a general expression of $\omega_{B}\tau$
    \begin{equation}
    \omega_{B}\tau\simeq(1-50)B_{0}/(10^{13}{\rm G}),
     \end{equation}
for young pulsars\,(Wang et al. 2020). It was found that when $B\leq 10^{15}$ G, the conductivity increases slightly with the increase in the magnetic field strength $B$ due to the quantum effects, the enhanced $B$ has a small effect on the matter in the low$-$density regions of the crust, and almost has no influence the matter in the high$-$density regions.
\subsection{Thermoplastic wave heating}
The thermoplastic instability launches a thermoplastic wave\,(TPW) analogous to a deflagration front. The propagation speed of the TPW is determined by $\chi$ and $\lambda$, and the characteristic width of the burning front $l$ is related to $\chi$ by
\begin{equation}
l\sim (\chi \Delta t)^{1/2}~~,
\end{equation}
where $\Delta t$ is the characteristic timescale of burning. The wave front velocity is given by
\begin{equation}
v\sim l/\Delta t\sim (\chi \mu_{B}/\lambda)^{1/2}~~.
\end{equation}
Considering an axisymmetric crustal plate in a cylindrical coordinate system $(r,\phi,z)$, the plate can be locally approximated by a slab model with $x=r$, $y=r\phi$ and $B_{y}=B_{\phi}$. The dissipation of the toroidal field $B_{\phi}$ causes the plate to rotate in the $\phi-$ direction.
\begin{figure}[htp!]
\vspace{0.12cm}
\centering
\includegraphics[width=0.5\textwidth]{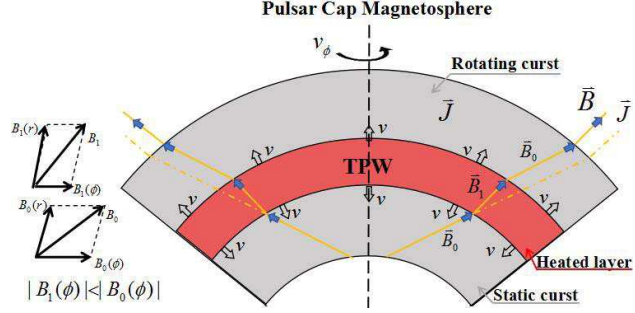}
\caption{ Crustal plate, that locates above the plastically heated layer, rotates very slowing with horizontal velocity $v_{\phi}$\,(Wang et al. 2019).}
\label{fig5}
\vspace{-0.12cm}
\end{figure}
In Fig.\,1, the blue arrows indicate the flow direction of the poloidal electric currents along the magnetic flux surfaces outside the radius of the light cylinder where the current closes in the next hemispheric surface. Dotted lines show the magnetic field lines before the pairs of TPW were launched. Once the pair of upward and downward fronts are launched from the plastically heated region, the upper crust rotates with respect to the static lower crust at a velocity
\begin{equation}
v_{\phi}=-4(b_{0}-1/\varsigma)v,
\end{equation}
where $b_{0}$ is the value of $b$ ($b=B_{\phi}/B_{z}=B_{\phi}/B_{r}$) at the onset of plastic flow and $\varsigma$ is a positive constant\,(Wang et al. 2019). Taking $r_{0}=R\sin\theta_{0}\sim 3.64$\,km and $b_{0}=9.3$ for the TPW structure, the heating rate by the TPW is estimated to be $\dot{q}\geq 1.6\times10^{21}$\,erg\,cm$^{-2}$\,s$^{-1}$ for PSR J1640$-$4631. The total energy dissipation rate is given by
\begin{eqnarray}
Q= \dot{q}_{\rm tot}\times\triangle V_{t}\approx \dot{q}_{\rm tot}\times 2\times(2\pi r_{0}\triangle r_{0}\triangle h),
\end{eqnarray}
where $\triangle V_{t}$ the total volume of the polar gaps in which the toroidal field dissipates, $\triangle r_{0}$ and $\triangle h$ are the polar ring width and height, respectively.  Then we have
\begin{equation}
L_{X}=\dot{q}_{\rm tot}V_{p}=S_{p}\delta T_{p}^{4}~,
\end{equation}
where $T_{p}$ is the predicted polar gap temperature, $\delta$ the Stefan-Boltzmann constant, and $S_{p}=4\pi R h_{p}$ the total polar cap area. The predicted X-ray flux with red shift in 2-10 keV band is given by
\begin{equation}
F_{X}^{\infty}=(\frac{R}{d_{12}})^{2}F_{X}(1-r_{g}/R)^{2},
\end{equation}
where $r_{g} = GM/c^{2}$ is the Schwarzschild radius, and $d_{12}$ is a distance in units of 12 kpc. Taking a density range $\rho\sim (6-9)\times10^{13}$\,g\,cm$^{-3}$, then we obtain  $F_{X}^{\infty}\sim (1.6-4.9)\times10^{-13}$ erg\,cm$^{-2}$\,s$^{-1}$, corresponding to $T_{p}\sim (1.8-2.4)\times10^{6}$\,K, which is very close to the observed soft X-ray flux\,(Gotthelf et al. 2014). Our results support the assumption that the observed soft X-ray in PSR J1640$-$4631 could originate from the TPW heating due to the dissipation of toroidal magnetic field near the polar gaps.
\section{The spin-down evolution }
\subsection{Braking indices of magnetars}
Due to electromagnetic radiation, particle winds, intense neutrino emission or gravitational radiation, a pulsar spins down. Allen et al. (1997) proposed that the long-term slowdown of a pulsar's rotation follows a power-law form $\tau_{\rm ext} \propto \Omega^{n}$ with $\tau_{\rm ext}$ being the external torque acting on the crust, $\Omega$ the rotational angular velocity, and $n$ the braking index, and gave an expression
\begin{equation}
	\dot{\Omega}=-K\Omega^{n},
\end{equation}
where $\dot{\Omega}$ is the derivative of $\Omega$, and $K$ is a positive parameter. The braking index of a pulsar is defined as
\begin{equation}
n=\frac{\Omega\ddot{\Omega}}{\dot{\Omega}^{2}}=\frac{\nu\ddot{\nu}}{\dot{\nu}^{2}}=2-\frac{P\ddot{P}}{\dot{P}^{2}},
\end{equation}
where $\ddot{\Omega}$ is the second derivative of $\Omega$, $\nu=\Omega/2\pi$ the spin frequency, and $P=1/\nu$ the spin period.

Because of lack of long-term pulsed emission in quiescence and strong timing noise, it is impossible to directly measure the braking index of a magnetar. Based on the estimated ages of their potentially associated SNRs and the timing parameters\,(Gao et al. 2016) measured the mean braking indices of eight magnetars with SNRs: SGR 0526$-$66, $n=2.40\pm0.04$; SGR 1627$-$41, $n=1.87\pm0.18$; PSR J1622$-$4950, $n=2.35\pm0.08$; CXOU J1714$-$3810, $n=2.1\pm0.9$; Swift J1834$-$0846, $n=1.08\pm0.04$; 1E 1841-045, $n=13\pm4$; SGR 0501$+$4516, $n=6.3\pm1.7$ and
1E 2259$+$586, $n=32\pm10$. We interpret the braking indices of $n<3$ within a combination of magneto-dipole radiation and wind aided braking, while the larger braking indices of $n>3$ for other three magnetars are attributed to the decay of external braking torque, which might be
caused by magnetic field decay.
\subsection{High-$n$ pulsar PSR J1640-4631}
The rotationally powered pulsar PSR J1640-4631 has
$\nu$=4.843 Hz and $\dot{\nu}=-2.28\times10^{-11}$\,Hz\,s$^{-1}$
(Archibald et al. 2016), implying a characteristic age of
$\tau_{c}=-\nu/2\dot{\nu}$=3370\,yr and surface dipole field $B_{p}=1.43\times10^{13}I^{1/2}_{45}$\,G, where $I_{45}$ is the
moment of inertia in units of $10^{45}$\,g\,cm$^{2}$\,(Gotthelf et al. 2014).
than 3, $n =3.15\pm0.03$, measured with
high precision.  In the case of a varying dipole magnetic field at the pole $B_{p}$, the braking index $n$ can be simply expressed as follows
\begin{equation}
n=3-4 \tau_{c} \frac{\dot{B}_{p}}{B_{p}}
\end{equation}
where $\dot{B}_{p}$ is the time derivative of $B_{p}$. For an increasing $B_{p}$,
we will always obtain $n<3$ owing to an increasing dipole braking torque, whereas $n>3$ for a decreasing $B_{p}$.
\begin{figure}[htp!]
\vspace{0.15cm}
\centering
\includegraphics[width=0.5\textwidth]{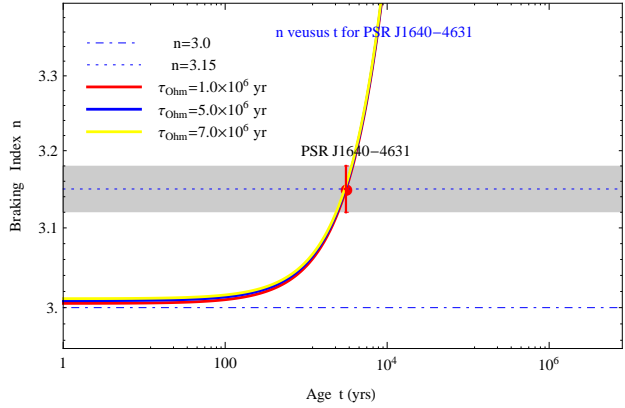}
\caption{ Braking index as a function of t for PSR J1640-4631. The measured value of $n$ is shown
with the red dot}
\label{fig5}
\vspace{-0.15cm}
\end{figure}
The evolution of the crustal magnetic field is phenomenologically divided into evolutionary stages: the initial stage with rapid decay, and a later stage with purely Ohmic dissipation. For simplicity, we adopted a phenomenological form of dipole magnetic filed, $B_{p}(t)$, given as Eq.(12) in Wang et al.\,(2020). After a complicated derivation, we obtained an analytic expression
\begin{eqnarray}
n&&=3+\frac{3 I c^{3}(\omega_{B} \tau)^{2} Z}{\pi^{2} R^{6} B_{0}^{2} \tau_{\rm Ohm}} \times\frac{a(a-e^{-Z t/
\tau_{Ohm}})}{e^{-2 Z t / \tau_{\rm Ohm}}}\times\nonumber\\
&& [P_{0}^{2}+\frac{4 \pi^{2} R^{6} B_{0}^{2} \tau_{\rm Ohm}}{3 I c^{3}(\omega_{B} \tau)^{2} Z}[\ln (a-1)+\frac{a}{a-1}
 -\frac{a}{a-e^{-Z t / \tau_{\rm Ohm}}}-\ln (a-e^{-Z t /\tau_{\rm Ohm}})]],
\end{eqnarray}
where $B_{0}$ is the initial value of $B_{p}$, $a=1+(\omega_{B}\tau)^{-1}$, and the red shift factor $Z=(1-2 G M/(c^{2}R))^{1/2}\approx 0.9$.
Combining Eq.\,(6) with Eq.\,(16), we obtained the relation of the braking index $n$ and time $t$ for PSR J1640$?$4631, as
shown in Fig.\,2. The solid red line represents the change trend expected by the dipole magnetic field decay model in the case of $\tau_{\rm Ohm}=
1.0\times10^{6}\,$yrs, $P_{0}=40.6$\,ms and $B_{0}=2.3752 \times10^{13}$\,G, while the solid blue line represents
the change trend expected by the dipole magnetic field decay model in the case of $\tau_{\rm Ohm}=5.0\times10^{6}$\,yrs,
 $P_{0}=39.9$\,ms and $B_{0}=2.3768 \times10^{13}\,$G, the  solid yellow line represents the change trend expected
 by the dipole magnetic field decay model in the case of $\tau_{\rm Ohm}=7.0\times10^{6}$\,yrs, $P_{0}=39.4$\,ms and
$B_{0}=2.3772 \times10^{13}$\,G. As can be seen from Fig.\,2, the braking index $n$ increases with the increase of $t$,
due to the decay of the dipole magnetic field. In Table 1, we list partial fitting values of Ohmic decay timescale, Hall drift timescale, magnetization parameter, initial dipole magnetic field, and initial rotational period of PSR J 1640-4631.
\begin{table}[htb!]
\caption{Fitted parameters of PSR J1640-4631.}
\centering
\begin{tabular}{ccccc}
\hline
$\tau_{\rm Ohm}$       &$\tau_{\rm Hall}$         &    $B_{0}$       &$\omega_{B}\tau$           &$P_{0}$   \\	
\hline
 yrs           &   yrs           &    G           &  $\frac{B_{0}}{10^{13}{\rm G}}$                &ms    \\
\hline
	1.0$\times10^{6}$    &8.45$\times10^{4}$    &2.3752$\times10^{13}$   &4.96    &40.6\\
    3.0$\times10^{6}$    &7.99$\times10^{4}$    &2.3758$\times10^{13}$   &15.8    &40.4\\
    5.0$\times10^{6}$    &7.89$\times10^{4}$    &2.3768$\times10^{13}$   &26.5    &39.9\\
    6.0$\times10^{6}$    &7.88$\times10^{4}$    &2.3769$\times10^{13}$   &31.9    &39.7\\
    7.0$\times10^{6}$    &7.86$\times10^{4}$    &2.3772$\times10^{13}$   &37.3    &39.4\\
    8.0$\times10^{6}$    &7.85$\times10^{4}$    &2.3774$\times10^{13}$   &42.7    &39.2\\
    9.0$\times10^{6}$    &7.84$\times10^{4}$    &2.3778$\times10^{13}$   &48.1    &38.9\\
    1.0$\times10^{7}$    &7.83$\times10^{4}$    &2.3782$\times10^{13}$   &53.5    &38.4\\
    2.0$\times10^{7}$    &7.80$\times10^{4}$    &2.3802$\times10^{13}$   &107.6   &38.3 \\
    3.0$\times10^{7}$    &7.78$\times10^{4}$    &2.3810$\times10^{13}$   &161.2   &37.8 \\
\hline
\end{tabular}

\end{table}
\section{Two works in progress}
\subsection{To explain anti-glitch}
Anti-glitch is referred to an unusual timing event, i.e., an abrupt and large decrease in $\nu\,(|\Delta \nu/\nu|\geq 10^{-7})$.  Archilbald et al.\,(2013) reported an anti-glitch observed in AXP 1E2259+586. Various models were proposed (see Huang \& Geng 2014, and references therein), but few of them took into account the magnetic flow model and the large braking index\,($n>3$) of this source. For example, if this anti-glitch was caused by strong particle wind\,(Kou et al. 2018), the braking index of 1E 2259+586 should be in a rather low range of $n<3$, which is not supported by our results\,(Gao et al. 2016). Alternatively, we suppose that 1E 2259+586 could have undergone an unusual decay of poloidal and toroidal magnetic fields, both of which maintain a stable elliptic configuration of the star. A long-term dipole poloidal field decay results in a decaying braking torque contributing to the high braking index. In order to maintain the balance of the star's configuration, the toroidal magnetic fields also decay.  In particular, when the decay of toroidal magnetic field accumulates to a certain extent, an abrupt stellar deformation starts to appear, companied with an decrease in ellipticity of the star, as shown in Fig.\,3, which results in an abrupt increase of $I$. According to the conservation of angular momentum, $J=I\Omega$, the star's spin frequency will decrease abruptly, dubbed the event an ''anti-glitch'' by astronomer.

To explain the magnetar anti-glitch, we will calculate the temperature-dependent shear rates as well as temperature-dependent
shear by considering magnetically driven plastic flows. The temperature-dependent shear rate $\dot{\Xi}$ is given as
\begin{equation}
\dot{\Xi}=2(\frac{V_{\rm act}}{V_{m}})(\frac{k_{B}T}{h}){\rm exp}(\frac{-\Delta E_{\rm act}}{k_{B}T}){\rm sinh(\frac{V_{\rm act}\tau}{2k_{B}T})},
\end{equation}
with $V_{act}$ being the activation volume, $V_{m}$ being the characteristic volume under consideration, $\Delta E_{\rm act}$ being a single energy barrier, $k_{B}$ the Boltzmann constant, $h$ the Plank constant, $\tau$ the local applied shearing stress, and $T$ is the temperature (Eyring \& Lee 1955). According to the thermal plastic flow model, the shear rate of the crust must be larger than the minimum
bound of 1 rad/year proposed by Beloborodov and Levin (2014). It is noted that a total strain $\delta\epsilon\sim 1$ corresponds to a
plastic energy release $>3\times10^{43}$\,erg (Thompson et al. 2017).  The magnetar bursts are are classed into three types primarily by energy released and their duration: short bursts with energies up to $10^{41}$\,er), intermediate flares with energies in the range $10^{41}-10^{43}$\,erg, and giant flares with energies in the range $10^{44}-10^{46}$\,erg. Within the magnetar scenario, the first kind of variability is thought to be driven by plastic deformations in the crust of magnetars which, in turn, induce changes in the magnetic
current configurations. During the anti-glitch events, the magnetars' bursts have always been observed.
\begin{figure}[htp!]
\vspace{0.125cm}
\centering
\includegraphics[width=0.5\textwidth]{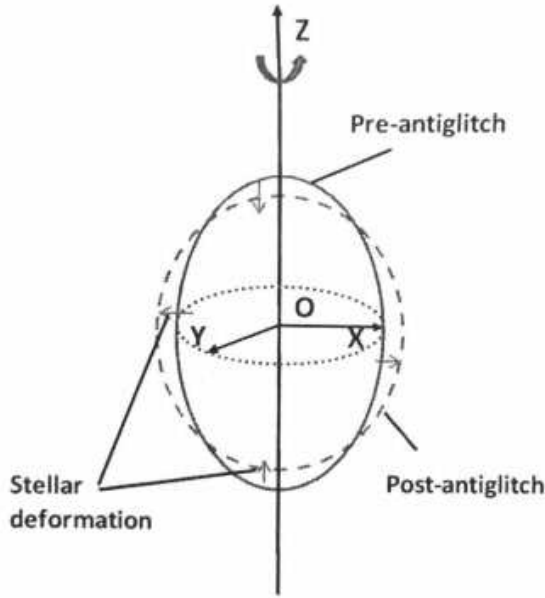}
\caption{ The diagram of  magnetar anti-glitch due to a sudden deformation with a decrease in ellipticity }
\label{fig5}
\vspace{-0.25cm}
\end{figure}
In addition, a likely ''anti-glitch'' event with $\Delta \nu/\nu\sim 5.8\times10^{-7}$ near MJD 54686 was also observed in 1E 1841-045\,(Mus et al. 2014), but there is no interpretation. Since 1E 1841$-$045 has a large braking index $n>3$, we suppose that the anti-glitch mechanism of 1E 1841$-$045 could be the same as that of 1E 2259+586. We will interpret the anti-glitch events of AXPs 1E 2259+586 and 1E 1841$-$045 magnetars by combining the timing and high-energy observations with the thermal-plastic-flow model, in which a temperature dependent shear rate, as well as plausible estimations of twist angle, temperature and magnetic field strengths, should be considered simultaneously. We surpose that, to judge whether a model for the magnetar anti-glitch is feasible, the braking index $n$ should be considered in the future. For an example, if an anti-glitch was observed on a magnetar, whose spin-down is dominated by particle wind, then its braking index should be less than 3, or more accurately, near 1.
\subsection{Revisiting pulsar braking index}
 Because of the influence of time noise and periodic jump, only a few of the pulsars can be measured well and the study of braking index is helpful to understand the evolution of pulsars. A dipole rotating in vacuum is subject to torque due to the MDR with spin-down and alignment components respectively given as
\begin{equation}
I\frac{d\Omega}{dt}=-\frac{2M^{2}\Omega^{3}}{3c^{3}}\sin^{2}\alpha,
\end{equation}
\begin{equation}
I\frac{d\alpha}{dt}=-\frac{2M^{2}\Omega^{2}}{3c^{3}}\sin\alpha\cos\alpha,
\end{equation}
where $\alpha$ the inclination angle between the rotation and magnetic axes, and $M=B_{d}R^{2}$ is the dipole magnetic moment\,(Davis \& Goldstein 1970) This model predicts a braking index of
\begin{equation}
	n=3+2\cot^{2}\alpha,
\end{equation}
which is always greater than 3, contradicting with the observed pulsars' braking indice\,(see Magalhaes eta l. 2012, 2016, Gao et al. 2017b for brief reviews), since the change of $M$ is ignored\,(Ek\c{s}i et al. 2016). The magnetic field decay occurring in pulsars\,(especially in magnetars) has been confirmed by a series of theoretical and observational studies\,(e.g., Kaspi \& Beloborodov 2017). The MDR model allows a change in magnetic moment $M$ and/or a change in the angle $\alpha$ of a pulsar, the latter is usually companied with a change in pulse profiles\,(Wen et al. 2016a, 2016b, 2020a, 2020b; Yan et al. 2019, 2020). If the changes of $M$ and $\alpha$ are considered simultaneously, the observed braking index $n_{\rm obs}$ is obtained as
\begin{equation}
n_{\rm obs}=3+\frac{2\Omega}{\dot{\Omega}}(\frac{\dot{\alpha}}{\tan\alpha}+\frac{\dot{M}}{M})
\end{equation}
from Eq.\,(18) and Eq.\,(19), where $\dot{\alpha}$ is the derivative of $\alpha$, and $\dot{M}$ the derivative of $M$. In order to achieve our goal,
we first convert Eq.\,(18) into the following form
\begin{equation}
-\frac{3}{2}Ic^{3}\frac{\dot{\Omega}}{\Omega^{3}}=M^{2}\sin^{2}\alpha,
\end{equation}
and then introduce the magnetic moment $M_{1}$, the inclination angle $\alpha_{1}$, the angular velocity $\Omega_{1}$ and its derivative $\dot{\Omega_{1}}$ at the time $t_{1}$, and the magnetic moment $M_{2}$, the inclination angle $\alpha_{2}$ the angular velocity $\Omega_{1}$ and its derivative $\dot{\Omega_{1}}$ at the time $t_{2}$. By introducing a sign of $A$ to denote the left-hand side of Eq.\,(22), we get
\begin{equation}
A_{1}=-\frac{3}{2}Ic^{3}\frac{\dot{\Omega_{1}}}{\Omega_{1}^{3}}=M_{1}^{2}\sin^{2}\alpha_{1},
\end{equation}
\begin{equation}
 A_{2}=-\frac{3}{2}Ic^{3}\frac{\dot{\Omega_{2}}}{\Omega_{2}^{3}}=M_{2}^{2}\sin^{2}\alpha_{2}
\end{equation}
for the first and the second measurements, respectively, where the change of moment of inertial $I$ is very small and can be ignored. We usually take the interval between the two measurements $\Delta t$ $(\Delta t=t_{2}-t_{1})$, as several years or decades. Correspondingly, $M_{2}=\Delta M+M_{1}$, $\alpha_{2}=\alpha_{1}+\Delta \alpha$, $\Omega_{2}=\Omega_{1}+\Delta \Omega$, and $\dot{\Omega_{2}}=\dot{\Omega_{1}}+\Delta\dot{\Omega}$ at the interval of $\Delta t$. From Eq.(23) and Eq.\,(24), it is natural
\begin{eqnarray}
 &&\sqrt{A_{2}}=M_{2}\sin\alpha_{2}=(M_{1}+\Delta M)\sin(\alpha_{1}+\Delta \alpha)\nonumber\\
 &&\approx M_{1}\sin\alpha_{1}+ \Delta M\sin\alpha_{1}+M_{1}\Delta \alpha\cos\alpha_{1},
\end{eqnarray}
where the second-order small quantity $\Delta M \Delta\alpha\cos\alpha_{1}$ is ignored.  Eq.(25) is converted into
\begin{equation}
\sqrt{A_{2}}-\sqrt{A_{1}}=\Delta M\sin\alpha_{1}+M_{1} \Delta\alpha\cos\alpha_{1}.
\end{equation}
Dividing both sides of Eq.\,(26) by $\sin\alpha_{1}\Delta t$, one have
\begin{equation}
\frac{\sqrt{A_{2}}-\sqrt{A_{1}}}{M_{1}\sin\alpha_{1}\Delta t}=\frac{\Delta M\sin\alpha_{1}}{M_{1}\sin\alpha_{1}\Delta t}+\frac{M_{1}\Delta \alpha\cos\alpha_{1}}{M_{1}\sin\alpha_{1}\Delta t}.
\end{equation}
Taking $\Delta M/\Delta t\sim dM_{1}/dt=\dot{M}_{1}$ and $\Delta \alpha/\Delta t\sim d\alpha_{1}/dt=\dot{\alpha}_{1}$, Eq.(27) becomes
\begin{equation}
\frac{1}{\Delta t}(\sqrt{\frac{A_{2}}{A_{1}}}-1)=\frac{\dot{M}_{1}}{M_{1}}+\frac{\dot{\alpha}_{1}}{\tan\alpha_{1}}.
\end{equation}
Then we  get $A_{2}/ A_{2}= \dot{\Omega}_{2}\Omega_{1}^{3}/(\dot{\Omega}_{1}\Omega_{2}^{3})=\dot{P}_{2}P_{2}/(\dot{P}_{1}P_{1})$, so Eq.\,(21) becomes
 \begin{eqnarray}
n_{\rm obs}=3+\frac{2\Omega_{1}}{\dot{\Omega_{1}}}(\sqrt{\frac{\dot{\Omega}_{2}\Omega_{1}^{3}}
{\dot{\Omega}_{1}\Omega_{2}^{3}}}-1)\frac{1}{\Delta t}
 =3+\frac{2\nu_{1}}{\dot{\nu_{1}}}(\sqrt{\frac{\dot{\nu}_{2}\nu_{1}^{3}}{\dot{\nu}_{1}\nu_{2}^{3}}}-1)\frac{1}{\Delta t},
\end{eqnarray}
where $\dot{\Omega}$ and $\dot{\nu}$ are always less than 0, if $\sqrt{\frac{\dot{\Omega}_{2}\Omega_{1}^{3}}{\dot{\Omega}_{1}\Omega_{2}^{3}}}> 1$, the second term on the right-hand side of the above equation is negative, then $n_{\rm obs}<3$, while $\sqrt{\frac{\dot{\Omega}_{2}\Omega_{1}^{3}}{\dot{\Omega}_{1}\Omega_{2}^{3}}}< 1$, the second term is positive and $n_{\rm obs}>3$.
Since Eq.\,(29) does not include the second derivative of $\Omega$ or $\nu$, it may have a wider range application than Eq.\,(14) in solving the pulsar braking index, more importantly, we can explain not only high braking index but also low braking index of pulsars using Eq.\,(29).  Compared to the original expression of Eq.\,(21),The advantage of Eq.\,(29) is that we do not need the observed values (or information) of magnetic moment $M$ and the angle $\alpha$. Alternatively, Eq.\,(29) is rewritten as
 \begin{eqnarray}
n_{\rm obs}=3-\frac{2P_{1}}{\dot{P_{1}}}(\sqrt{\frac{\dot{P}_{2}P_{2}}{\dot{P}_{1}P_{1}}}-1)\frac{1}{\Delta t},
\end{eqnarray}
where $\dot{P}$ is always larger than 0, if $(\sqrt{\frac{\dot{P}_{2}P_{2}}{\dot{P}_{1}P_{1}}}-1)<0$, the second term on the right-hand side of Eq.(30) is positive, then $n_{\rm obs}>3$, while $(\sqrt{\frac{\dot{P}_{2}P_{2}}{\dot{P}_{1}P_{1}}}-1)> 0$, and $n_{\rm obs}<3$. Due to frequent glitches and strong timing noise, it is difficult to obtain an accurate braking index of a pulsar\,(such as a magnetar), although the second derivatives of its angular velocity $\Omega$, spin frequency $\nu$, and spin period $P$ exist. In the future, we will calculate braking indices of more young pulsars using Eq.\,(29) or Eq.\,(30), and give plausible interpretations in the coupled model of the decay of dipole magnetic moment and the change of inclination angle\,(a possible increase or decrease in $\alpha$).
\section{ Summary and outlook}
We have reviewed our recent works on braking index, crustal conductivity,magnetization parameter, magnetic field evolution, and temperature-dependent plastic flows in young and strongly magnetized pulsars. It seems to us that a deeper X-ray observation together with a detailed study of future activity periods of magnetars, including simultaneous X-ray/optical monitoring, could not only verify our present results but also shed light onto its detailed nature and discern whether the remote source is an ultra-compact low-mass X-ray binary or an isolated neutron star displaying a new manifestation of magnetar activity. In the future, we will further investigate the pulsars' braking indices using our model.  Note that a series of observed transient phenomena, e.g., X-ray bursts, superbursts, magnetar flares and glitches in the spin-down evolution are thought to possibly originate in the crust of magnetars.

\section*{Acknowledgments}
This work is partly supported by Natural Science Foundation of Xinjiang, China under grant number 2018D01A24






%


\end{document}